\documentclass[fleqn,usenatbib]{mnras}

\usepackage{newtxtext,newtxmath}

\usepackage[T1]{fontenc}

\DeclareRobustCommand{\VAN}[3]{#2}
\let\VANthebibliography\thebibliography
\def\thebibliography{\DeclareRobustCommand{\VAN}[3]{##3}\VANthebibliography}

\usepackage{graphicx}	
\usepackage{amsmath}	

\usepackage{graphicx}
 \usepackage{setspace}
\usepackage{textcomp}     
\usepackage{url}
\usepackage[]{amsmath}
\usepackage{amsmath}
\usepackage{nccmath}
\usepackage{lineno}

\usepackage{hyperref}
\usepackage[flushleft]{threeparttable}
\newcommand{\ti}{{3I/ATLAS }}
\newcommand{\tins}{{3I/ATLAS}}
\newcommand{\psj}{{Planetary Science Journal}}

\title[Interstellar comet 3I/ATLAS]{Interstellar comet 3I/ATLAS: discovery and physical description}

\author[B. T. Bolin et al.]{Bryce T. Bolin,$^{1,}$\thanks{These authors contributed equally to this work.}\thanks{Email: bbolin@eurekasci.com}
Matthew Belyakov,$^{2}$\footnotemark[1]
Christoffer Fremling,$^{3}$\footnotemark[1]
Matthew J. Graham,$^{4}$
\newauthor
Ahmed. M. Abdelaziz$^{5}$
Eslam Elhosseiny,$^{5}$
Candace L. Gray,$^{6}$
Carl Ingebretsen,$^{7}$
\newauthor
Gracyn Jewett,$^{8}$
Carey M. Lisse,$^{9}$
Sergey Karpov,$^{10}$
Mukremin Kilic,$^{8}$
Martin Ma\v{s}ek,$^{10}$
\newauthor
Mona Molham,$^{5}$
Diana Roderick,$^{3}$
Ali Takey,$^{5}$
Laura-May Abron,$^{11}$
\newauthor
Michael W. Coughlin,$^{12}$
Cheng-Han Hsieh,$^{13}$\thanks{NASA Hubble Fellowship Sagan Fellow}
Keith S. Noll,$^{14}$
Ian Wong$^{15}$
\\
$^{1}$Eureka Scientific, Oakland, CA 94602, USA,\\
$^{2}$Division of Geological and Planetary Sciences, California Institute of Technology, Pasadena, CA 91125, USA,\\
$^{3}$Caltech Optical Observatories, California Institute of Technology, Pasadena, CA 91125, USA,\\
$^{4}$Cahill Center for Astrophysics, California Institute of Technology, Pasadena, CA, 91125, USA,\\
$^{5}$National Research Institute of Astronomy and Geophysics (NRIAG), 1 El-marsad St., 11421 Helwan, Cairo, Egypt\\
$^{6}$Department of Astronomy, Apache Point Observatory, New Mexico State University, Las Cruces, NM, USA,\\
$^{7}$Department of Physics and Astronomy, Johns Hopkins University, Baltimore, MD 21218, USA,\\
$^{8}$Homer L. Dodge Department of Physics and Astronomy, University of Oklahoma, Norman, OK 73019, USA,\\
$^{9}$Johns Hopkins University Applied Physics Laboratory, Laurel, MD 20723, USA,\\
$^{10}$FZU - Institute of Physics of the Czech Academy of Sciences, Na Slovance 1999/2, CZ-182 21, Praha, Czech Republic,\\
$^{11}$Griffith Observatory, Los Angeles, CA 90027, USA,\\
$^{12}$School of Physics and Astronomy, University of Minnesota, Minneapolis, Minnesota 55455, USA\\
$^{13}$Department of Astronomy, University of Texas at Austin, Austin, TX 21218, USA,\\
$^{14}$Goddard Space Flight Center, Greenbelt, MD 20771, USA,\\
$^{15}$Space Telescope Science Institute, 3700 San Martin Drive, Baltimore, MD 21218, MD
}
\date{Accepted XXX. Received YYY; in original form ZZZ}
\pubyear{2025}
\begin{document}
\label{firstpage}
\pagerange{\pageref{firstpage}--\pageref{lastpage}}
\maketitle
\begin{abstract}
We present the characteristics of interstellar comet \tins, discovered on 2025 July 1 by the Asteroid Terrestrial-impact Last Alert System. The comet has eccentricity $\simeq$ 6.08 and velocity at infinity $\simeq$ 57 km/s, indicating an interstellar origin. We obtained B,V, R, I, g, r, i, and z photometry at Kottamia Astronomical Observatory, Palomar Observatory, and Apache Point Observatory on 2025 July 2, 3, and 6. We measured colour indices B--V=0.98$\pm$0.23, V--R=0.71$\pm$0.09, R--I=0.14$\pm$0.10, g--r=0.84$\pm$0.05, r--i=0.16$\pm$0.03, i--z=--0.02$\pm$0.07, g--i=1.00$\pm$0.05, and a spectral slope of 16.0$\pm$1.9 $\%$/100 nm. We estimate the comet's dust cross-section within 10,000 km to be 184.6$\pm$4.6 km$^2$, assuming a 0.1 albedo. \tins's coma has FWHM$\simeq$2.2 arcsec and A(0$^\circ$)f$\rho$=280.8$\pm$3.2 cm. We estimate \ti is ejecting \textmu m to mm dust at $\sim$0.01-1 m/s, with a mass loss of $\sim$0.1 - 1.0 kg/s.
\end{abstract}



\section{Introduction}

Interstellar objects (ISO) are small bodies that originate from outside the solar system whose orbital trajectories intersect the solar system \citep[for a review, see][]{JewittISOReview2024}. Present-day studies of extra-solar debris disks are limited to their micron and mm-sized components \citep[][]{Su2024,Xie2025}, but ISOs provide opportunities for the study of macroscopic bodies that originated from other stellar systems \citep[e.g.,][]{Bolin2018,Hopkins2023}. Until recently, only two known interstellar objects, 1I/`Oumuamua and 2I/Borisov have been found \citep[][]{JewittISOReview2024}. 1I/'Oumuamua appeared inactive but was subsequently  found to have significant non-gravitational acceleration \citep[][]{Micheli2018}, while 2I/Borisov showed clear evidence of an extended coma \citep[][]{Guzik2019a,Kim20202I,Bolin2020HST} and volatile compounds \citep[][]{Bodewits2020,Cordiner2020}.

The third known macroscopic interstellar object, comet \tins, was discovered on 2025 July 1 \citep[][]{Denneau20253I} by the Asteroid Terrestrial-impact Last Alert System (ATLAS). This paper describes its discovery, and its subsequent imaging characterisation by the Kottamia Astronomical Observatory (KAO) 1.88-m telescope, the Palomar 200-inch telescope, and the Apache Point Observatory (APO) Astrophysical Research Consortium 3.5-m (ARC 3.5-,m) telescope. We use the techniques described by \citet[][]{Bolin20202I,Bolin2024E3} for characterizing active objects with multi-band photometry to test the coma-centric colours of \tins. In addition, we apply the surface brightness profile analysis techniques of \citet[][]{Bolin2021LD2, Bolin2025KY26} to characterize the activity of \ti and its dust properties.

\section{Observations}

\begin{figure}
\centering
\includegraphics[scale=0.37]{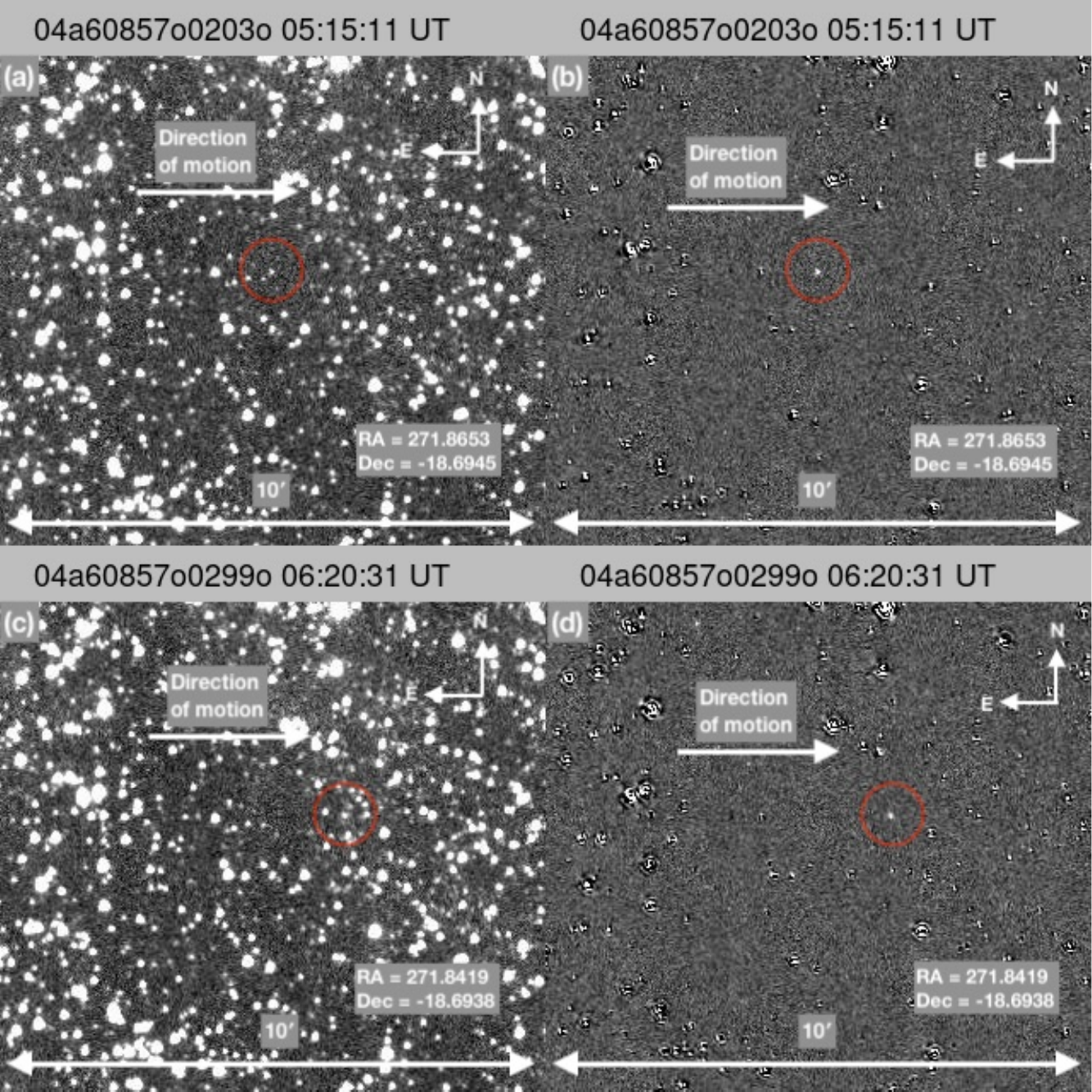}
\caption{\textbf{Cutout images from the first and fourth discovery observations of \ti from the ATLAS facility in Chile, spanning approximately one hour.} \ti is moving at 73.31 arcsec/h against the stellar background. The cardinal directions and direction of motion are indicated with arrows, and \ti identified within the red circle. (a, b) Unsubtracted and subtracted images from 05:15:11 UT; (c, d) Unsubtracted and subtracted images from 06:20:31 UT. Image credit: Larry Denneau / ATLAS / University of Hawaii / NASA.}
\label{figdiscovery}
\end{figure}

Initially designated as A11pl3Z, interstellar comet \ti was discovered at the ATLAS Rio Hurtado, Chile facility (Minor Planet Center (MPC) observing code W68). The initial discovery observations of \ti were made on 2025 July 1 05:15:11 in the ATLAS o-band filter as seen in Panels a-d of Figure \ref{figdiscovery} \citep[][]{Denneau20253I}. The comet was discovered at right ascension, declination of 18:07:27.68 and -18:41:40.2 and was detected in four o-band exposures taken between 2025 July 1 05:15:11and 2025 July 1 06:20:31. A list containing details of the ATLAS observations is presented in Section \ref{obsdet} and Table \ref{t:obs} of the Supplemental Material.

For the follow up and characterization of \tins, used the Kottamia Astronomical Observatory 1.88-m telescope's (KAO 1.88-m's) Kottamia Faint Imaging Spectro-Polarimeter (KFISP), the Next Generation Palomar Spectrograph (NGPS) instrument on the Palomar 200-inch telescope, and the Apache Point Observatory Astrophysical Research Consortium 3.5-m (ARC 3.5-m)/ Astrophysical Research Consortium Telescope Imaging Camera (ARCTIC) on 2025 July 2, 2025 July 3, and 2025 July 6. Imaging in B,V,R, and I, was taken with KFISP, g, r, and i images were taken with the NGPS, and g, r, i, and z images were taken with ARCTIC. Spectra of \ti were also taken with P200 and ARC 3.5-m, which will be described in a forthcoming paper \citep[][]{Belyakov2025ATel3I}.

\begin{figure}
\centering
\includegraphics[scale=0.31]{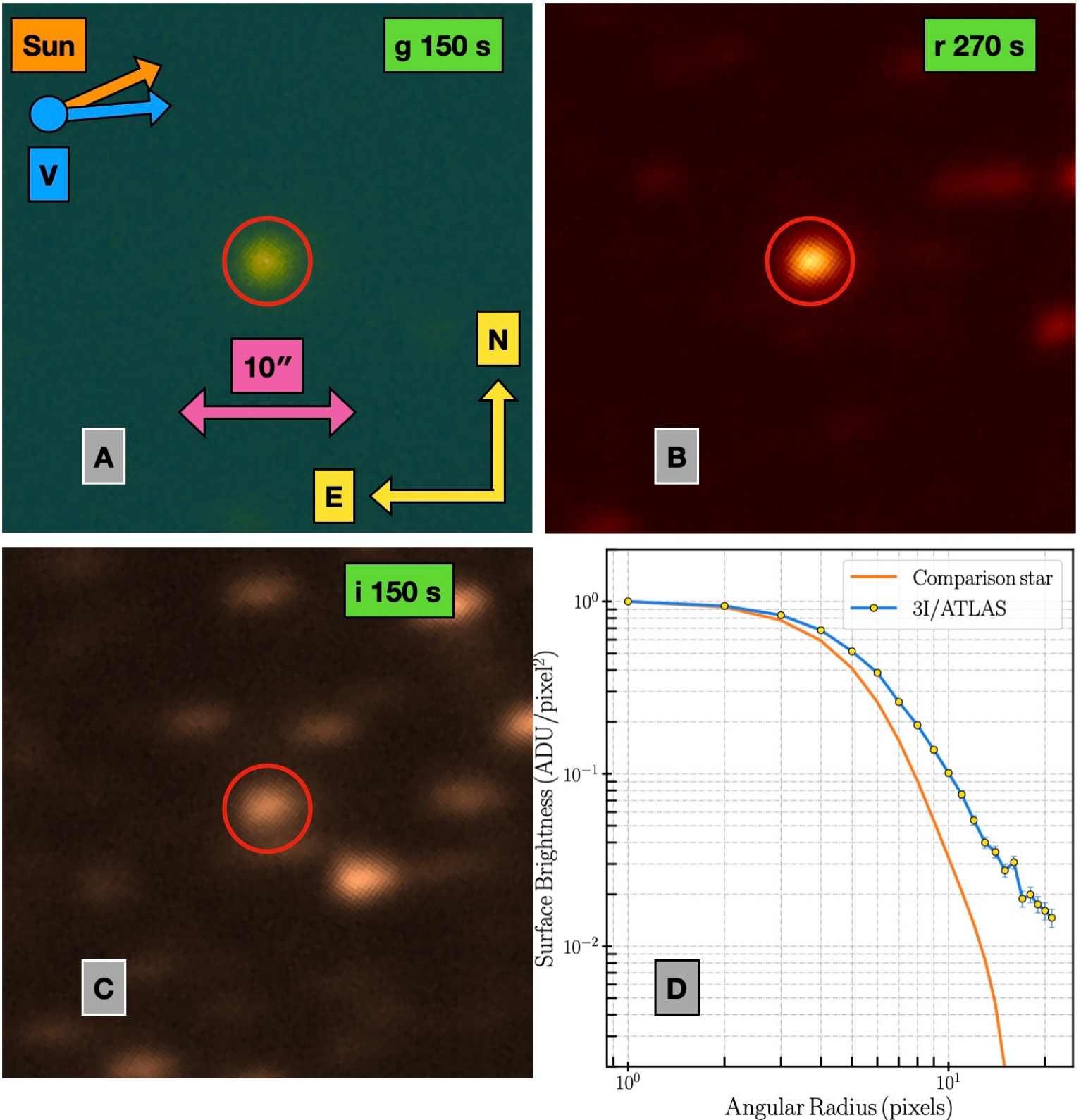}
\caption{\textbf{Deep g, r, and i-band imaging of \ti taken with NGPS on 2025 July 3.} \textbf{Panel A:} a composite stack of \ti in two g-band images with a total exposure time of 150 s. The comet is indicated with a circle. An arrow indicating a width of 10\arcsec~is shown for scale. The cardinal, solar, and orbital motion directions are indicated. \textbf{Panel B:} a combined stack of \ti in r-band of three r-band images with a total exposure time of 270 s. \textbf{Panel C:} a combined stack of two i-band images of \ti with a cumulative exposure time of 150 s. Streaks are due to incompletely removed background star trails. \textbf{Panel D:} the azimuthally-averaged surface brightness profile of the combined r-band stack plotted as yellow data points with error bars. The error bars are smaller than the marker for the first twelve data points. A nearby comparison star's azimuthally-averaged surface brightness profile is over-plotted as an orange line.}
\label{figprof}
\end{figure}

\begin{figure}
\centering
\includegraphics[scale=0.26]{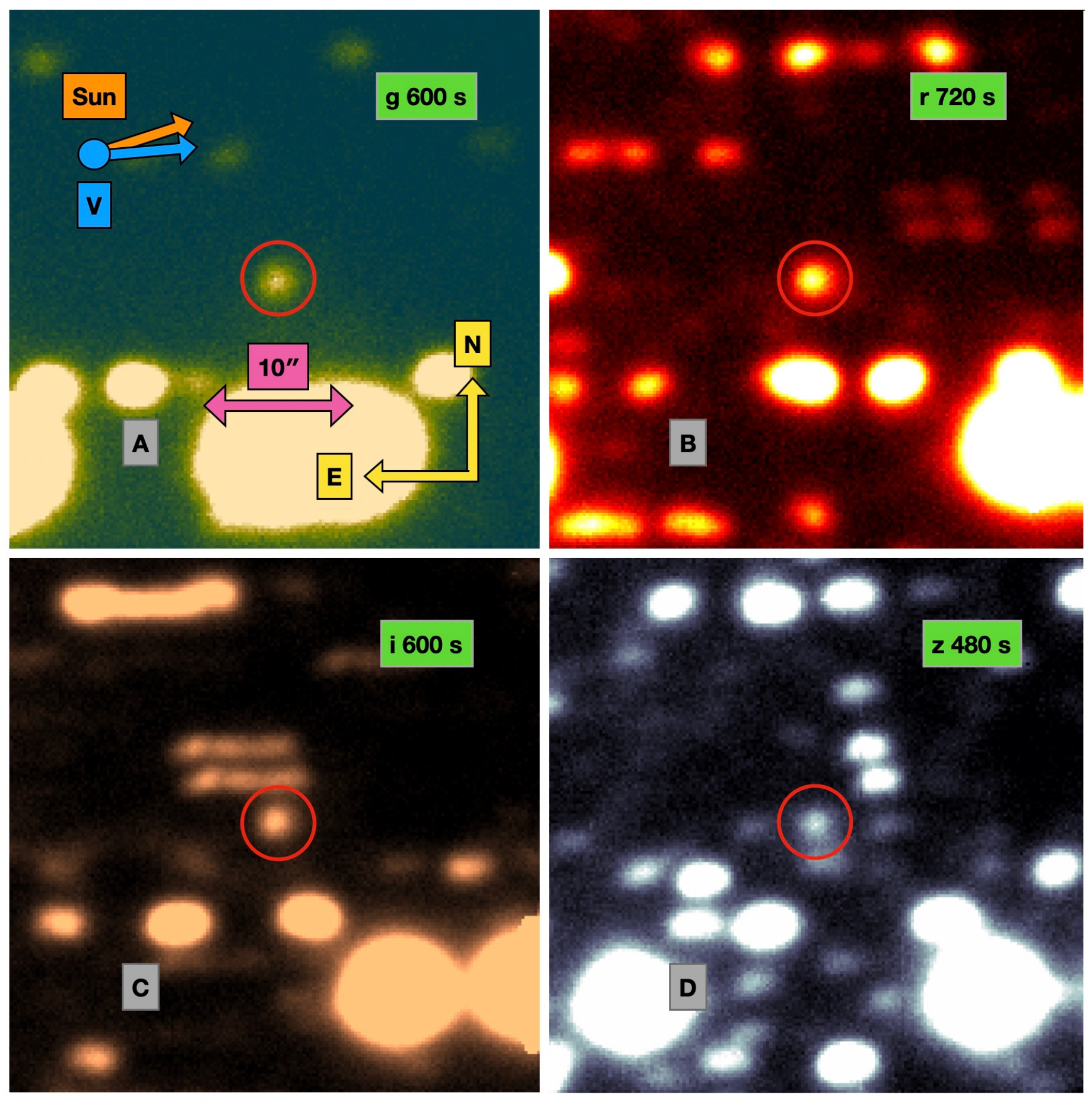}
\caption{\textbf{Imaging g, r, and i, and z-band stacks of \ti taken with ARCTIC on 2025 July 6.} \textbf{Panel A:} a median-combined stack of 5 x 120 s g filter images of \ti encircled in red. \textbf{Panel B:} a median-combined stack of 6 x 120 s r filter images of \tins. \textbf{Panel C:} a median-combined stack of 5 x 120 s i filter images of \tins. \textbf{Panel D:} a median-combined stack of 4 x 120 s z filter images of \tins. The images were taken with the telescope tracked at the asteroid's apparent rate of motion. The cardinal, solar, and orbital motion directions are indicated, and a double arrow 10 arcsec wide is included for scale.}
\label{figapo}
\end{figure}

Images from all three facilities were reduced using biases and flats taken for each of the filters. Image stacks of the NGPS and ARCTIC data are shown in the mosaic images in Figs. \ref{figprof} and \ref{figapo}. The multi-band data were calibrated using the Pan-STARRS1 catalog \citep[][]{Chambers2016} and the colour transformations from \citep[][]{Tonry2012} to convert Pan-STARR1 photometry magnitudes into SDSS photometry magnitudes. The photometry of \ti was measured using a 4\arcsec~radius circular aperture (10,000 km from the comet's nucleus at its 3.36 - 3.45 au distance from the Earth during our KAO 1.88-m, P200, and ARCTIC 3.5-m observaitons).  We used a 6.5-7.8\arcsec~radius sky background annulus to measure our photomery. The seeing at the KAO 1.88-m, the P200, and the ARC 3.5-m measured between 1.3-1.6 arcsec in images containing the comet. A list of the details of the KAO 1.88-m, P200, and ARC 3.5-m observations is presented in Section \ref{obsdet} and Table \ref{t:obs} of the Supplemental Material.

\section{Results}

\subsection{Orbit}
\begin{figure}
\centering
\includegraphics[scale=0.16]{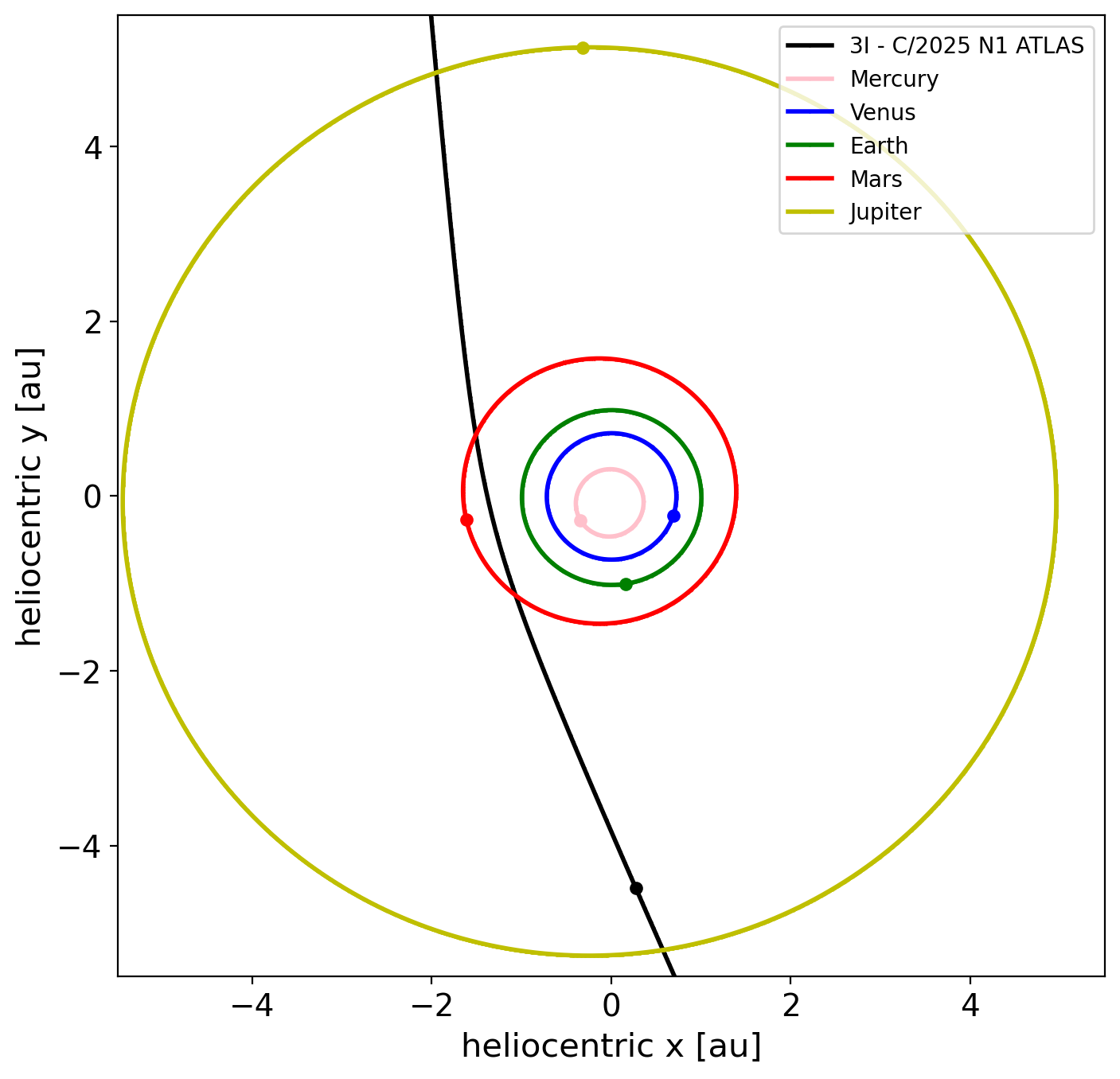}
\caption{\textbf{Orbits of \ti and solar system planets.} \textbf{Top panel:} Ecliptic pole view of the orbits and positions of \ti (black), Mercury (pink), Venus (blue), Earth (green), Mars (red) and Jupiter (khaki) at the time of the discovery of \ti on 2025 July 1 05:15:11. Positions of the planets and \ti are from JPL HORIZONS.}
\label{figorbit}
\end{figure}

The astrometry of \ti from ATLAS taken in the discovery images on 2025 July 1 was reported to the MPC. Astrometry of the comet in $\sim$319 observations taken from 97 observational facilities around the world between 2025 June 14.25 UTC and 2025 July 4 UTC was subsequently reported to the MPC \footnote{\url{https://minorplanetcenter.net/db_search/show_object?utf8=\%E2\%9C\%93&object_id=3I}, accessed on 2025 July 4.}. The orbital solution from JPL HORIZONS, using observations taken between June 14, 2025, and July 4, 2025, is shown in Table \ref{tabel}. The location and orbit of \ti at the time of its discovery on 2025 July 1 is shown on Figure \ref{figorbit} based on values from JPL Horizons\footnote{\url{https://ssd.jpl.nasa.gov/horizons/app.html\#/}}.

\subsection{Visible photometry}

\begin{figure}
\centering
\includegraphics[scale=.27]{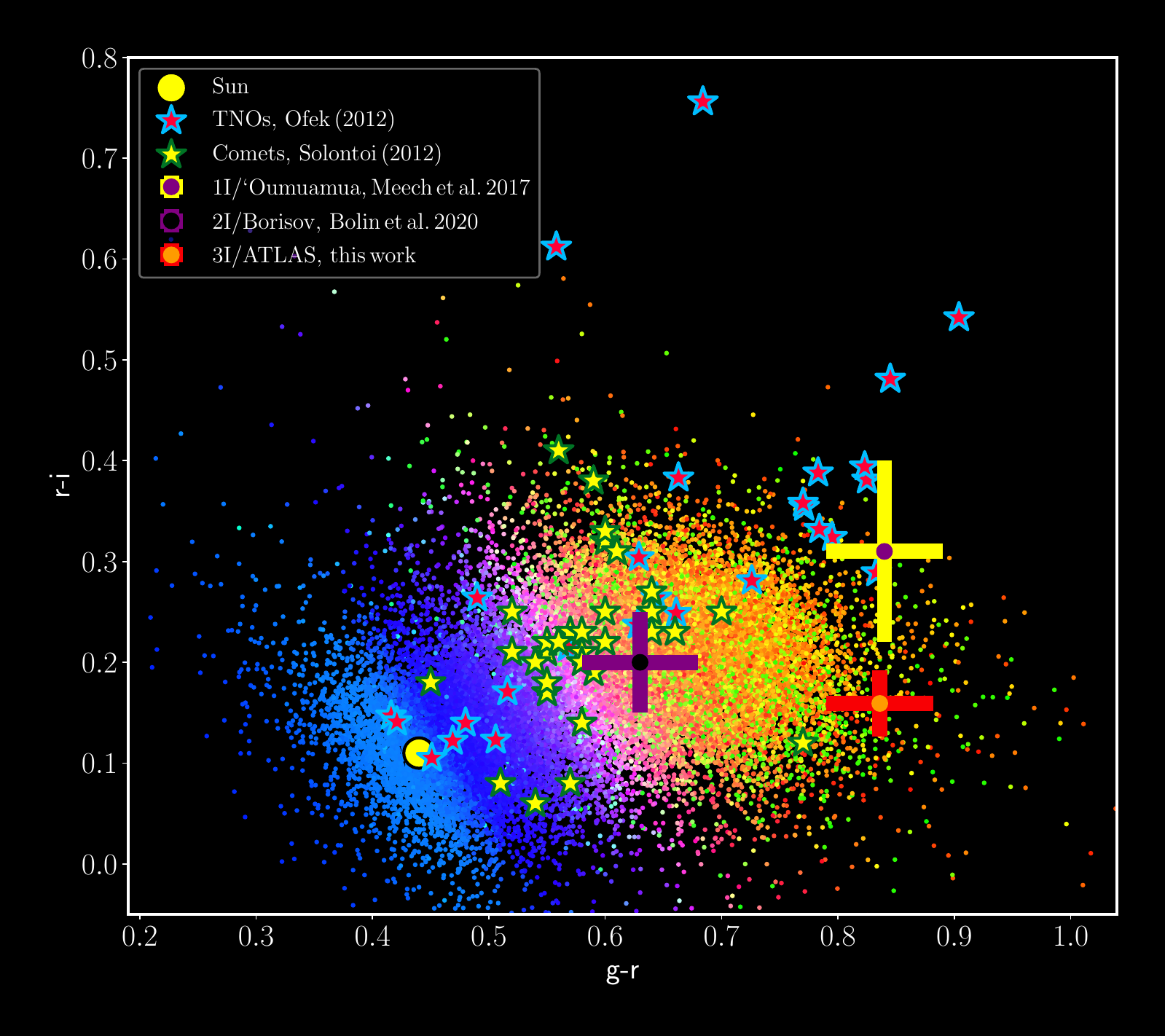}
\includegraphics[scale=.27]{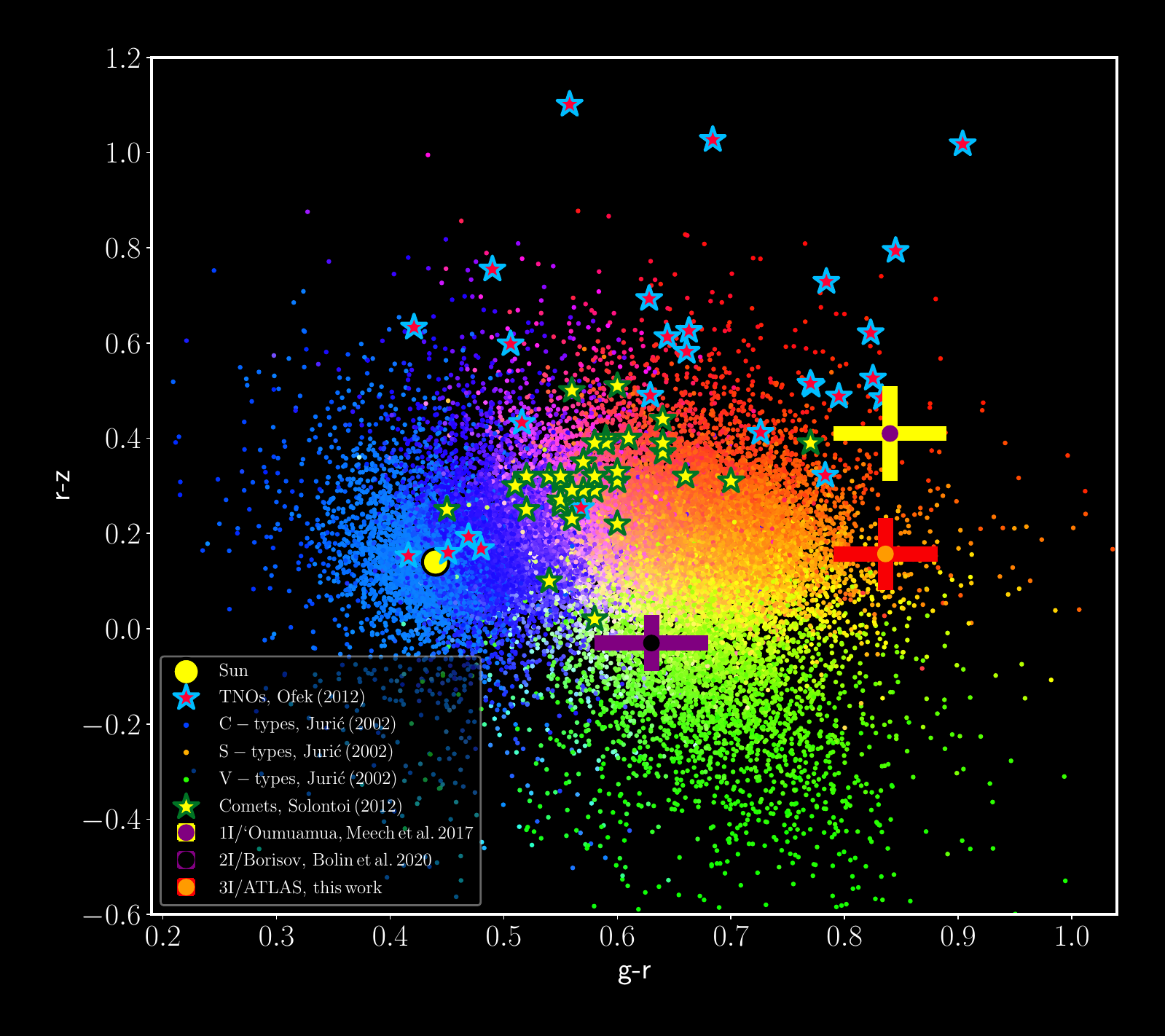}
\caption{\textbf{The g, r, i, and z colours of \ti from our measurements compared with solar system objects and other ISOs.} \textbf{Top panel:} The g--r vs. r--i colors of \ti and other objects. We use the scheme of \citet[][]{Ivezic2002} where blue dots correspond to C-complex asteroids, red dots correspond to S-Complex asteroids, and green dots correspond to V-complex asteroids. We also plot the colours of active comets from \citet[][]{Solontoi2012} and trans-Neptunian objects from \citet[][]{Ofek2012}. \textbf{Bottom panel:}  the same as in the top panel, but showing the g-r vs. r-z colours of \ti and other small bodies. In the top panel, the g--r and r--i colors from the NGPS and ARCTIC observations have been averaged together. In the bottom panel, only the g--r color from the NGPS and ARCTIC have been averaged, and the r--z color for ARCTIC has been plotted.}
\label{figcolor}
\end{figure}

We computed the brightness of \ti in the KAO 1.88-m images of B=18.90$\pm$ 0.21, V=17.92 $\pm$ 0.01, R=17.22 $\pm$ 0.05, and I=17.08 $\pm$ 0.09, in the NGPS images of g=18.72$\pm$0.02, r=17.86$\pm$0.01, and i=17.72$\pm$0.02, and g=18.58$\pm$0.06, r=17.77$\pm$0.02, i=17.59$\pm$0.04, and z=17.61$\pm$0.05 in the ARCTIC images. We used the NGPS r-band stack to search for evidence of an extended coma in \tins. The surface brightness profile extracted from the r-band images of \ti (blue line in Panel D of Figure \ref{figprof}) shows evidence of being extended, having $>$2-10 times the flux compared to the surface brightness profile of a comparison star (orange line in the same panel) outside of 1.3 arcsec of both profiles. The FWHM of the coma in the NGPS r-band stack is $\sim$2.2 arcsec compared to the 1.6 arcsec FWHM of background stars in the sidereal-tracked images. \ti's colours are B--V=0.98$\pm$0.23, V--R=0.71$\pm$0.09, R--I=0.14$\pm$0.10, g--r=0.84$\pm$0.05 mag, r--i=0.16$\pm$0.03 mag, i--z=--0.02$\pm$0.07 mag, and g--i=1.00$\pm$0.05 mag corresponding to a spectral slope of 16.0$\pm$1.9 $\%$/100 nm., somewhat more red than interstellar comet 2I/Borisov which had g--r=0.63$\pm$0.05, r--i=0.20$\pm$0.02, g--i=0.83$\pm$0.05, and a spectral slope of 9.7$\pm$1.8 $\%$/100 nm. The colors of \ti derived from our observations are similar to those reported by other photometric and spectroscopic studies \citep[e.g.,][]{Opitom2025}. We plot the g--r, r--i, and r--z colours of \ti in Figure \ref{figcolor}.

We use the g, r, and i photometry obtained by the P200/NGPS to calculate the A(0$^\circ$)f$\rho$  parameter, a proxy for dust production rate normalised to a phase angle of 0 degrees \citep[][]{AHearn1984, Schleicher1998, Marcus2007}. We find g-band A(0$^\circ$)f$\rho$ = 187.7$\pm$4.1 cm, r-band A(0$^\circ$)f$\rho$ = 280.8$\pm$3.1 cm, i-band A(0$^\circ$)f$\rho$ = 287.8$\pm$4.5 cm, and z-band A(0$^\circ$)f$\rho$ = 289.3$\pm$13.7 cm, typical values for active solar system comets \citep[][]{AHearn1995,Kelley2013}. We calculate the absolute magnitude, $H$, according to the following equation \citep[][]{Jewitt1991}:
\begin{equation}
\label{eqn.brightness}
H = V - 2.5 \mathrm{log_{10}}(r_h \Delta) + \Phi(\alpha)
\end{equation}
where $V$ is the apparent V magnitude (determined from our g and r observations to be $V$ = 18.28$\pm$0.03),  $r_h$ is the heliocentric distance of the comet on 2025 July 3 of 4.43 au, $\Delta$ is the distance of the comet from the Earth on 2025 July 3 of 3.43 au, $\Phi(\alpha)$ is a function describing the brightening of the comet which we approximate with $\Phi(\alpha)$ = $--$0.04$\alpha$, appropriate for comets observed at phase angles with $<$20 degrees \citep[][]{Bertini2017}, and $\alpha$ is the phase angle of the comet in degrees, of 2.65 degrees. Using Eq.~1, we obtain $H$=12.27$\pm$0.03 mag, cautioning that the uncertainties of $H$ inferred from Eq.~1 are systematically underestimated due to the presently unknown phase function for \tins. We translate the $H$ magnitude computed from Eq.~1 into an effective cross-section, $C$, in units of km$^2$ within a 10,000 km aperture using 
\begin{equation}
C = 1.5 \times 10^{6}\, p_v^{-1} \, 10^{-0.4H}
\end{equation}
from \citet[][]{Jewitt2016}, where $p_v$ is the albedo of the comet, assumed to be 0.10, typical for comet dust \citep[][]{Jewitt1986, Kolokolova2004,Zubko2017}, finding a value of 184.6$\pm$4.6 km$^2$. 

We can use the size of the coma measured in the sunward direction to provide an estimate of the dust ejection velocity, $v_e$, \citep[][]{Jewitt1987}:
\begin{equation}
v_e = \frac{\left[2D_t\beta g_{\odot}\right]^{\frac{1}{2}}}{r_h[\mathrm{au}]}
\end{equation}
where D$_t$ is the extent of the coma in the sunward direction measured in km, $\beta$ is a quantity inversely proportional to the particle size, $a_s$, in \textmu m, $g_{\odot}$ = 0.006 m s$^{-2}$, and $r_h$ is the heliocentric distance in au. We calculate D$_t$ using \citep[][]{Hsieh2021QN173}:
\begin{equation}
D_t \propto (\theta_{\mathrm{3I}}^2 - \theta_{\mathrm{star}}^2)^{\frac{1}{2}}
\end{equation}
where $\theta_{\mathrm{3I}}$ is the half-width half maximum measured from our r-band stack of 1.1 arcsec, $\theta_{\mathrm{star}}$ is the measured half-width half maximum of background stars of $\sim$0.8 arcsec, resulting in a $D_t$ of 0.7 arcsec, which translates to 1850 km at the 3.43 au distance of the comet. Plugging our calculated value of $D_t$ of 1850 km into Eq.~4 gives $v_e$$\sim$1.1$\beta^\frac{1}{2}$m s$^{-1}$ suggesting that \text m-scale to mm-scale particles are ejected from the comet at $\sim$0.01-1 m/s. The dust mass loss, $\dot{M}$ can be roughly estimated from $v_e$, A(0$^\circ$)f$\rho$, the dust albedo $p_v$, the radius of the dust particles, $r_d$, and the dust density, $\rho_d$, with the following relation derived from \citep[][]{Fink2012}:
\begin{equation}
\dot{M} = \frac{4 \pi \rho_d v_e r_d A(0^\circ)f\rho}{3 p_v} 
\end{equation}
plugging in our values of $v_e$ = $\sim$10$^{-2}$-1 m/s for \textmu m to mm-sized particles,  A(0$^\circ$)f$\rho$ = 281 cm, $\rho_d$ = 750 kg m$^{-3}$ \citep[][]{Fulle2016}, and $p_v$ = 0.1 results in $\dot{M}$ = $\sim$0.1 - 1.0 kg/s, similar to interstellar comet 2I/Borisov when it was on its inbound trajectory and located 2.7 au from the Sun \citep{Fitzsimmons2019,Jewitt2019}, and similar to some Jupiter Family Comets in the solar system \citep[][]{Gillan2024}.

\section{Discussion and conclusion}
Interstellar comet \ti exhibits an active appearance and colours similar to some solar system comets, as seen in Figures \ref{figprof} and \ref{figcolor}. However, it seems to be significantly more red in its color compared to the other interstellar comet 2I (g--r=0.63$\pm$0.05, r--i=0.20$\pm$0.05, g--i=0.83$\pm$0.05, and r--z=--0.03$\pm$0.06), however, is less red compared to 1I/'Oumuamua (g--r=0.84$\pm$0.05, r--i=0.31$\pm$0.09, g--i=1.15$\pm$0.10, and r--z=0.41$\pm$0.10). However, this latter comparison may be inaccurate due to the fact that 3I/ATLAS is an active body, and 1I/'Oumuamua was only very weakly active as evidenced by the visible lack of a coma \citep[][]{Micheli2018}. A complete table of the physical properties of the different interstellar objects is included in Table \ref{t:objs}. Interstellar comet 3I/ATLAS appears to be more red than the majority of comets as seen in Fig.~4. The increased redness of 3I compared to 2I and other active comets may be due to the nature of its activity if it lacks the emission of gas molecules present in the g-band wavelength such as CN, C$_2$, and C$_2$ \citep[][]{Farnham2000}. With our small sample size of three objects, it appears that the interstellar objects span a wide range in color. Additionally, 3I shows evidence of having a similar dust mass loss compared to 2I, but has a much higher dust mass loss rate than 1I, which was observed to show a negligible amount of dust mass loss.

While it is difficult to speculate on the size of 3I due to the coma obscuring its nucleus, future high-resolution observations will provide a better test of its size than ground-based, seeing-limited observations can. Additional future observations of \ti in the visible will test its activity and potentially its evolution as it passes through perihelion in 2025 October at a distance 1.35 au from the Sun. Observations covering the near-infrared and submillimeter wavelengths will enable the study of its ice and dust grain properties. Space-based observations spanning the near-infrared to mid-infrared will facilitate a test of the production of cometary gases in its coma and activity-driving mechanism. Dedicated and serendipitous observations by surveys such as the Rubin Observatory Legacy Survey of Space and Time \citep[][]{Ivezic2019,Bolin2025TS,Dorsey2025ISORubin} will provide coverage of distant comets and ISOs which may lead to a greater understanding of the extrasolar planetesimal population \citep[][]{Hopkins2023}.

\section*{Acknowledgements}
The authors wish to acknowledge Larry Denneau for sharing the discovery images of \ti and the details of the discovery observations. The authors also wish to thank Robert Jedicke for help with the orbital diagram, and for the scientific discussion of the paper, and Yanga. R. Fern\'{a}ndez for providing comments to the manuscript. Our work is based on observations taken at Palomar Observatory. The authors wish to thank the Palomar observing staff for their help with the observations. The authors also wish to recognize and acknowledge the cultural significance that Palomar Mountain has for the Pauma Band of the Luise\~{n}o Indians. MK and GJ acknowledge support from the NSF under grant AST-2205736 and the NASA under grants 80NSSC22K0479, 80NSSC24K0380, and 80NSSC24K0436. MWC acknowledges support from the National Science Foundation with grant numbers PHY-3117998, PHY-2308862, and PHY-2409481. MJG acknowledges support from the National Science Foundation with grant numbers PHY-2117997. The National Research Institute of Astronomy and Geophysics team acknowledges financial support from the Egyptian Science, Technology $\&$ Innovation Funding Authority (STDF) under grant number 48102. Based on observations obtained with the Apache Point Observatory 3.5-meter telescope, which is owned and operated by the Astrophysical Research Consortium. MM thanks the projects MEYS LM2023032, LM2023047.

\section*{Data Availability}
The data underlying this article will be shared on reasonable request to the corresponding author. The data for the Palomar and Apache Point observations can be found at the following repository \citep[][]{Belyakov20253IDataCaltech}.

\section*{Supplemental Material}
The Supplemental Material for this manuscript is available online.



\bibliographystyle{mnras}
\bibliography{/Users/bolin/Dropbox/Projects/latex_references_commands_bib_styles/neobib} 

\begin{thebibliography}{}
\makeatletter
\relax
\def\mn@urlcharsother{\let\do\@makeother \do\$\do\&\do\#\do\^\do\_\do\%\do\~}
\def\mn@doi{\begingroup\mn@urlcharsother \@ifnextchar [ {\mn@doi@}
  {\mn@doi@[]}}
\def\mn@doi@[#1]#2{\def\@tempa{#1}\ifx\@tempa\@empty \href
  {http://dx.doi.org/#2} {doi:#2}\else \href {http://dx.doi.org/#2} {#1}\fi
  \endgroup}
\def\mn@eprint#1#2{\mn@eprint@#1:#2::\@nil}
\def\mn@eprint@arXiv#1{\href {http://arxiv.org/abs/#1} {{\tt arXiv:#1}}}
\def\mn@eprint@dblp#1{\href {http://dblp.uni-trier.de/rec/bibtex/#1.xml}
  {dblp:#1}}
\def\mn@eprint@#1:#2:#3:#4\@nil{\def\@tempa {#1}\def\@tempb {#2}\def\@tempc
  {#3}\ifx \@tempc \@empty \let \@tempc \@tempb \let \@tempb \@tempa \fi \ifx
  \@tempb \@empty \def\@tempb {arXiv}\fi \@ifundefined
  {mn@eprint@\@tempb}{\@tempb:\@tempc}{\expandafter \expandafter \csname
  mn@eprint@\@tempb\endcsname \expandafter{\@tempc}}}

\bibitem[\protect\citeauthoryear{{A'Hearn}, {Schleicher}, {Millis}, {Feldman}
  \& {Thompson}}{{A'Hearn} et~al.}{1984}]{AHearn1984}
{A'Hearn} M.~F.,  {Schleicher} D.~G.,  {Millis} R.~L.,  {Feldman} P.~D.,
  {Thompson} D.~T.,  1984, \mn@doi [\aj] {10.1086/113552}, \href
  {https://ui.adsabs.harvard.edu/abs/1984AJ.....89..579A} {89, 579}

\bibitem[\protect\citeauthoryear{{A'Hearn}, {Millis}, {Schleicher}, {Osip}  \&
  {Birch}}{{A'Hearn} et~al.}{1995}]{AHearn1995}
{A'Hearn} M.~F.,  {Millis} R.~C.,  {Schleicher} D.~O.,  {Osip} D.~J.,   {Birch}
  P.~V.,  1995, \mn@doi [\icarus] {10.1006/icar.1995.1190}, \href
  {https://ui.adsabs.harvard.edu/abs/1995Icar..118..223A} {118, 223}

\bibitem[\protect\citeauthoryear{{Azzam} et~al.,}{{Azzam}
  et~al.}{2022}]{Azzam2022}
{Azzam} Y.~A.,  et~al., 2022, \mn@doi [Experimental Astronomy]
  {10.1007/s10686-021-09802-z}, \href
  {https://ui.adsabs.harvard.edu/abs/2022ExA....53...45A} {53, 45}

\bibitem[\protect\citeauthoryear{Belyakov, Fremling, Graham, Mukremin, Bolin,
  Jewett  \& Lisse}{Belyakov et~al.}{2025a}]{Belyakov20253IDataCaltech}
Belyakov M.,  Fremling C.,  Graham M.~J.,  Mukremin K.,  Bolin B.,  Jewett G.,
   Lisse C.,  2025a, Palomar P200 and APO Imaging and Spectroscopy of
  Interstellar Object 3I/Atlas, \mn@doi{10.22002/qdce4-pvm83}

\bibitem[\protect\citeauthoryear{{Belyakov}, {Bolin}, {Fremling}  \&
  {Graham}}{{Belyakov} et~al.}{2025b}]{Belyakov2025ATel3I}
{Belyakov} M.,  {Bolin} B.~T.,  {Fremling} C.,   {Graham} M.,  2025b, The
  Astronomer's Telegram, \href
  {https://ui.adsabs.harvard.edu/abs/2025ATel17276....1B} {17276, 1}

\bibitem[\protect\citeauthoryear{{Bertini} et~al.,}{{Bertini}
  et~al.}{2017}]{Bertini2017}
{Bertini} I.,  et~al., 2017, \mn@doi [\mnras] {10.1093/mnras/stx1850}, \href
  {https://ui.adsabs.harvard.edu/abs/2017MNRAS.469S.404B} {469, S404}

\bibitem[\protect\citeauthoryear{{Bodewits} et~al.,}{{Bodewits}
  et~al.}{2020}]{Bodewits2020}
{Bodewits} D.,  et~al., 2020, \mn@doi [Nature Astronomy]
  {10.1038/s41550-020-1095-2}, \href
  {https://ui.adsabs.harvard.edu/abs/2020NatAs.tmp...85B} {}

\bibitem[\protect\citeauthoryear{{Bolin} \& {Lisse}}{{Bolin} \&
  {Lisse}}{2020}]{Bolin2020HST}
{Bolin} B.~T.,  {Lisse} C.~M.,  2020, \mn@doi [\mnras]
  {10.1093/mnras/staa2192}, \href
  {https://ui.adsabs.harvard.edu/abs/2020MNRAS.497.4031B} {497, 4031}

\bibitem[\protect\citeauthoryear{{Bolin} et~al.,}{{Bolin}
  et~al.}{2018}]{Bolin2018}
{Bolin} B.~T.,  et~al., 2018, \mn@doi [\apjl] {10.3847/2041-8213/aaa0c9}, \href
  {http://adsabs.harvard.edu/abs/2018ApJ...852L...2B} {852, L2}

\bibitem[\protect\citeauthoryear{{Bolin} et~al.,}{{Bolin}
  et~al.}{2020}]{Bolin20202I}
{Bolin} B.~T.,  et~al., 2020, \mn@doi [\aj] {10.3847/1538-3881/ab9305}, \href
  {https://ui.adsabs.harvard.edu/abs/2020AJ....160...26B} {160, 26}

\bibitem[\protect\citeauthoryear{{Bolin} et~al.,}{{Bolin}
  et~al.}{2021}]{Bolin2021LD2}
{Bolin} B.~T.,  et~al., 2021, \mn@doi [\aj] {10.3847/1538-3881/abd94b}, \href
  {https://ui.adsabs.harvard.edu/abs/2021AJ....161..116B} {161, 116}

\bibitem[\protect\citeauthoryear{{Bolin} et~al.,}{{Bolin}
  et~al.}{2024}]{Bolin2024E3}
{Bolin} B.~T.,  et~al., 2024, \mn@doi [\mnras] {10.1093/mnrasl/slad139}, \href
  {https://ui.adsabs.harvard.edu/abs/2024MNRAS.527L..42B} {527, L42}

\bibitem[\protect\citeauthoryear{{Bolin} et~al.,}{{Bolin}
  et~al.}{2025a}]{Bolin2025KY26}
{Bolin} B.~T.,  et~al., 2025a, \mn@doi [\aj] {10.3847/1538-3881/adccbe}, \href
  {https://ui.adsabs.harvard.edu/abs/2025AJ....169..303B} {169, 303}

\bibitem[\protect\citeauthoryear{{Bolin} et~al.,}{{Bolin}
  et~al.}{2025b}]{Bolin2025TS}
{Bolin} B.~T.,  et~al., 2025b, \mn@doi [\icarus]
  {10.1016/j.icarus.2024.116333}, \href
  {https://ui.adsabs.harvard.edu/abs/2025Icar..42516333B} {425, 116333}

\bibitem[\protect\citeauthoryear{{Chambers} et~al.,}{{Chambers}
  et~al.}{2016}]{Chambers2016}
{Chambers} K.~C.,  et~al., 2016, preprint, \href
  {http://adsabs.harvard.edu/abs/2016arXiv161205560C} {} (\mn@eprint {arXiv}
  {1612.05560})

\bibitem[\protect\citeauthoryear{{Cordiner} et~al.,}{{Cordiner}
  et~al.}{2020}]{Cordiner2020}
{Cordiner} M.~A.,  et~al., 2020, \mn@doi [Nature Astronomy]
  {10.1038/s41550-020-1087-2}, \href
  {https://ui.adsabs.harvard.edu/abs/2020NatAs...4..861C} {4, 861}

\bibitem[\protect\citeauthoryear{{Cousins}}{{Cousins}}{1976}]{Cousins1976}
{Cousins} A.~W.~J.,  1976, \memras, \href
  {https://ui.adsabs.harvard.edu/abs/1976MmRAS..81...25C} {81, 25}

\bibitem[\protect\citeauthoryear{{Denneau}, {Siverd}, {Tonry}, {Erasmus},
  {Fitzsimmons}  \& {Robinson}}{{Denneau} et~al.}{2025}]{Denneau20253I}
{Denneau} L.,  {Siverd} R.,  {Tonry} J.,  {Erasmus} N.,  {Fitzsimmons} A.,
  {Robinson} J.,  2025, MPEC

\bibitem[\protect\citeauthoryear{{Dorsey}, {Hopkins}, {Bannister}, {Lawler},
  {Lintott}, {Parker}  \& {Forbes}}{{Dorsey} et~al.}{2025}]{Dorsey2025ISORubin}
{Dorsey} R.~C.,  {Hopkins} M.~J.,  {Bannister} M.~T.,  {Lawler} S.~M.,
  {Lintott} C.,  {Parker} A.~H.,   {Forbes} J.~C.,  2025, \mn@doi [arXiv
  e-prints] {10.48550/arXiv.2502.16741}, \href
  {https://ui.adsabs.harvard.edu/abs/2025arXiv250216741D} {p. arXiv:2502.16741}

\bibitem[\protect\citeauthoryear{{Farnham}, {Schleicher}  \&
  {A'Hearn}}{{Farnham} et~al.}{2000}]{Farnham2000}
{Farnham} T.~L.,  {Schleicher} D.~G.,   {A'Hearn} M.~F.,  2000, \mn@doi
  [\icarus] {10.1006/icar.2000.6420}, \href
  {https://ui.adsabs.harvard.edu/abs/2000Icar..147..180F} {147, 180}

\bibitem[\protect\citeauthoryear{{Fink} \& {Rubin}}{{Fink} \&
  {Rubin}}{2012}]{Fink2012}
{Fink} U.,  {Rubin} M.,  2012, \mn@doi [\icarus]
  {10.1016/j.icarus.2012.09.001}, \href
  {https://ui.adsabs.harvard.edu/abs/2012Icar..221..721F} {221, 721}

\bibitem[\protect\citeauthoryear{{Fitzsimmons} et~al.,}{{Fitzsimmons}
  et~al.}{2019}]{Fitzsimmons2019}
{Fitzsimmons} A.,  et~al., 2019, \mn@doi [\apjl] {10.3847/2041-8213/ab49fc},
  \href {https://ui.adsabs.harvard.edu/abs/2019ApJ...885L...9F} {885, L9}

\bibitem[\protect\citeauthoryear{{Fukugita}, {Ichikawa}, {Gunn}, {Doi},
  {Shimasaku}  \& {Schneider}}{{Fukugita} et~al.}{1996}]{Fukugita1996}
{Fukugita} M.,  {Ichikawa} T.,  {Gunn} J.~E.,  {Doi} M.,  {Shimasaku} K.,
  {Schneider} D.~P.,  1996, \mn@doi [\aj] {10.1086/117915}, \href
  {https://ui.adsabs.harvard.edu/abs/1996AJ....111.1748F} {111, 1748}

\bibitem[\protect\citeauthoryear{{Fulle} et~al.,}{{Fulle}
  et~al.}{2016}]{Fulle2016}
{Fulle} M.,  et~al., 2016, \mn@doi [\mnras] {10.1093/mnras/stw2299}, \href
  {https://ui.adsabs.harvard.edu/abs/2016MNRAS.462S.132F} {462, S132}

\bibitem[\protect\citeauthoryear{{Gillan}, {Fitzsimmons}, {Denneau}, {Siverd},
  {Smith}, {Tonry}  \& {Young}}{{Gillan} et~al.}{2024}]{Gillan2024}
{Gillan} A.~F.,  {Fitzsimmons} A.,  {Denneau} L.,  {Siverd} R.~J.,  {Smith}
  K.~W.,  {Tonry} J.~L.,   {Young} D.~R.,  2024, \mn@doi [\psj]
  {10.3847/PSJ/ad1394}, \href
  {https://ui.adsabs.harvard.edu/abs/2024PSJ.....5...25G} {5, 25}

\bibitem[\protect\citeauthoryear{{Gunn} et~al.,}{{Gunn}
  et~al.}{1998}]{Gunn1998}
{Gunn} J.~E.,  et~al., 1998, \mn@doi [\aj] {10.1086/300645}, \href
  {https://ui.adsabs.harvard.edu/abs/1998AJ....116.3040G} {116, 3040}

\bibitem[\protect\citeauthoryear{{Guzik}, {Drahus}, {Rusek}, {Waniak},
  {Cannizzaro}  \& {Pastor-Marazuela}}{{Guzik} et~al.}{2019}]{Guzik2019a}
{Guzik} P.,  {Drahus} M.,  {Rusek} K.,  {Waniak} W.,  {Cannizzaro} G.,
  {Pastor-Marazuela} I.,  2019, \mn@doi [Nature Astronomy]
  {10.1038/s41550-019-0931-8}, \href
  {https://ui.adsabs.harvard.edu/abs/2019NatAs.tmp..467G} {p.~467}

\bibitem[\protect\citeauthoryear{{Hopkins}, {Lintott}, {Bannister}, {Mackereth}
   \& {Forbes}}{{Hopkins} et~al.}{2023}]{Hopkins2023}
{Hopkins} M.~J.,  {Lintott} C.,  {Bannister} M.~T.,  {Mackereth} J.~T.,
  {Forbes} J.~C.,  2023, \mn@doi [\aj] {10.3847/1538-3881/ad03e6}, \href
  {https://ui.adsabs.harvard.edu/abs/2023AJ....166..241H} {166, 241}

\bibitem[\protect\citeauthoryear{{Hsieh} et~al.,}{{Hsieh}
  et~al.}{2021}]{Hsieh2021QN173}
{Hsieh} H.~H.,  et~al., 2021, \mn@doi [\apjl] {10.3847/2041-8213/ac2c62}, \href
  {https://ui.adsabs.harvard.edu/abs/2021ApJ...922L...9H} {922, L9}

\bibitem[\protect\citeauthoryear{{Huehnerhoff} et~al.,}{{Huehnerhoff}
  et~al.}{2016}]{Huehnerhoff2016}
{Huehnerhoff} J.,  et~al., 2016, in Ground-based and Airborne Instrumentation
  for Astronomy VI. p. 99085H, \mn@doi{10.1117/12.2234214}

\bibitem[\protect\citeauthoryear{{Ivezi{\'c}} et~al.,}{{Ivezi{\'c}}
  et~al.}{2002}]{Ivezic2002}
{Ivezi{\'c}} {\v Z}.,  et~al., 2002, \mn@doi [\aj] {10.1086/344077}, \href
  {http://adsabs.harvard.edu/abs/2002AJ....124.2943I} {124, 2943}

\bibitem[\protect\citeauthoryear{{Ivezi{\'c}} et~al.,}{{Ivezi{\'c}}
  et~al.}{2019}]{Ivezic2019}
{Ivezi{\'c}} {\v{Z}}.,  et~al., 2019, \mn@doi [\apj]
  {10.3847/1538-4357/ab042c}, \href
  {https://ui.adsabs.harvard.edu/abs/2019ApJ...873..111I} {873, 111}

\bibitem[\protect\citeauthoryear{{Jewitt}}{{Jewitt}}{1991}]{Jewitt1991}
{Jewitt} D.,  1991, in {Newburn} Jr. R.~L.,  {Neugebauer} M.,   {Rahe} J.,
  eds,  Astrophysics and Space Science Library Vol. 167, IAU Colloq. 116:
  Comets in the post-Halley era. pp 19--65,
  \mn@doi{10.1007/978-94-011-3378-4_2}

\bibitem[\protect\citeauthoryear{{Jewitt}}{{Jewitt}}{2024}]{JewittISOReview2024}
{Jewitt} D.,  2024, \mn@doi [arXiv e-prints] {10.48550/arXiv.2407.06475}, \href
  {https://ui.adsabs.harvard.edu/abs/2024arXiv240706475J} {p. arXiv:2407.06475}

\bibitem[\protect\citeauthoryear{{Jewitt} \& {Luu}}{{Jewitt} \&
  {Luu}}{2019}]{Jewitt2019}
{Jewitt} D.,  {Luu} J.,  2019, \mn@doi [\apjl] {10.3847/2041-8213/ab530b},
  \href {https://ui.adsabs.harvard.edu/abs/2019ApJ...886L..29J} {886, L29}

\bibitem[\protect\citeauthoryear{{Jewitt} \& {Meech}}{{Jewitt} \&
  {Meech}}{1986}]{Jewitt1986}
{Jewitt} D.,  {Meech} K.~J.,  1986, \mn@doi [\apj] {10.1086/164745}, \href
  {https://ui.adsabs.harvard.edu/abs/1986ApJ...310..937J} {310, 937}

\bibitem[\protect\citeauthoryear{{Jewitt} \& {Meech}}{{Jewitt} \&
  {Meech}}{1987}]{Jewitt1987}
{Jewitt} D.~C.,  {Meech} K.~J.,  1987, \mn@doi [\apj] {10.1086/165347}, \href
  {https://ui.adsabs.harvard.edu/abs/1987ApJ...317..992J} {317, 992}

\bibitem[\protect\citeauthoryear{{Jewitt} et~al.,}{{Jewitt}
  et~al.}{2016}]{Jewitt2016}
{Jewitt} D.,  et~al., 2016, \mn@doi [\apjl] {10.3847/2041-8205/829/1/L8}, \href
  {https://ui.adsabs.harvard.edu/abs/2016ApJ...829L...8J} {829, L8}

\bibitem[\protect\citeauthoryear{{Kasliwal} et~al.,}{{Kasliwal}
  et~al.}{2024}]{Kasliwal2024}
{Kasliwal} M.~M.,  et~al., 2024, Transient Name Server AstroNote, \href
  {https://ui.adsabs.harvard.edu/abs/2024TNSAN.340....1K} {340, 1}

\bibitem[\protect\citeauthoryear{{Kelley} et~al.,}{{Kelley}
  et~al.}{2013}]{Kelley2013}
{Kelley} M.~S.,  et~al., 2013, \mn@doi [\icarus]
  {10.1016/j.icarus.2013.04.012}, \href
  {https://ui.adsabs.harvard.edu/abs/2013Icar..225..475K} {225, 475}

\bibitem[\protect\citeauthoryear{{Kim}, {Jewitt}, {Mutchler}, {Agarwal}, {Hui}
  \& {Weaver}}{{Kim} et~al.}{2020}]{Kim20202I}
{Kim} Y.,  {Jewitt} D.,  {Mutchler} M.,  {Agarwal} J.,  {Hui} M.-T.,   {Weaver}
  H.,  2020, \mn@doi [\apjl] {10.3847/2041-8213/ab9228}, \href
  {https://ui.adsabs.harvard.edu/abs/2020ApJ...895L..34K} {895, L34}

\bibitem[\protect\citeauthoryear{{Kolokolova}, {Hanner}, {Levasseur-Regourd}
  \& {Gustafson}}{{Kolokolova} et~al.}{2004}]{Kolokolova2004}
{Kolokolova} L.,  {Hanner} M.~S.,  {Levasseur-Regourd} A.-C.,   {Gustafson}
  B.~{\AA}.~S.,  2004, {Physical properties of cometary dust from light
  scattering and thermal emission}.
pp 577--604

\bibitem[\protect\citeauthoryear{{Marcus}}{{Marcus}}{2007}]{Marcus2007}
{Marcus} J.~N.,  2007, International Comet Quarterly, \href
  {https://ui.adsabs.harvard.edu/abs/2007ICQ....29...39M} {29, 39}

\bibitem[\protect\citeauthoryear{{Meech} et~al.,}{{Meech}
  et~al.}{2017}]{Meech2017}
{Meech} K.~J.,  et~al., 2017, \mn@doi [\nat] {10.1038/nature25020}, \href
  {https://ui.adsabs.harvard.edu/abs/2017Natur.552..378M} {552, 378}

\bibitem[\protect\citeauthoryear{{Micheli} et~al.,}{{Micheli}
  et~al.}{2018}]{Micheli2018}
{Micheli} M.,  et~al., 2018, \mn@doi [\nat] {10.1038/s41586-018-0254-4}, \href
  {https://ui.adsabs.harvard.edu/abs/2018Natur.559..223M} {559, 223}

\bibitem[\protect\citeauthoryear{Ofek}{Ofek}{2012}]{Ofek2012}
Ofek E.~O.,  2012, The Astrophysical Journal, 749, 10

\bibitem[\protect\citeauthoryear{{Opitom} et~al.,}{{Opitom}
  et~al.}{2025}]{Opitom2025}
{Opitom} C.,  et~al., 2025, \mn@doi [arXiv e-prints]
  {10.48550/arXiv.2507.05226}, \href
  {https://ui.adsabs.harvard.edu/abs/2025arXiv250705226O} {p. arXiv:2507.05226}

\bibitem[\protect\citeauthoryear{{Osman}}{{Osman}}{2001}]{Osman2001}
{Osman} A. M.~I.,  2001, in IAU General Assembly. pp 179--186

\bibitem[\protect\citeauthoryear{{Schleicher}, {Millis}  \&
  {Birch}}{{Schleicher} et~al.}{1998}]{Schleicher1998}
{Schleicher} D.~G.,  {Millis} R.~L.,   {Birch} P.~V.,  1998, \mn@doi [\icarus]
  {10.1006/icar.1997.5902}, \href
  {https://ui.adsabs.harvard.edu/abs/1998Icar..132..397S} {132, 397}

\bibitem[\protect\citeauthoryear{{Solontoi} et~al.,}{{Solontoi}
  et~al.}{2012}]{Solontoi2012}
{Solontoi} M.,  et~al., 2012, \mn@doi [\icarus] {10.1016/j.icarus.2011.10.008},
  \href {http://adsabs.harvard.edu/abs/2012Icar..218..571S} {218, 571}

\bibitem[\protect\citeauthoryear{{Su} et~al.,}{{Su} et~al.}{2024}]{Su2024}
{Su} K. Y.~L.,  et~al., 2024, \mn@doi [\apj] {10.3847/1538-4357/ad8cde}, \href
  {https://ui.adsabs.harvard.edu/abs/2024ApJ...977..277S} {977, 277}

\bibitem[\protect\citeauthoryear{{Tang}, {Hu}  \& {Ji}}{{Tang}
  et~al.}{2023}]{Tang2023}
{Tang} L.,  {Hu} Z.,   {Ji} H.,  2023, in {Busse} L.~E.,  {Soskind} Y.,  eds,
  Society of Photo-Optical Instrumentation Engineers (SPIE) Conference Series
  Vol. 12428, Digital Optical Technologies 2023. p. 1242816,
  \mn@doi{10.1117/12.2657248}

\bibitem[\protect\citeauthoryear{{Tonry} et~al.,}{{Tonry}
  et~al.}{2012}]{Tonry2012}
{Tonry} J.~L.,  et~al., 2012, \mn@doi [\apj] {10.1088/0004-637X/750/2/99},
  \href {http://adsabs.harvard.edu/abs/2012ApJ...750...99T} {750, 99}

\bibitem[\protect\citeauthoryear{{Tonry} et~al.,}{{Tonry}
  et~al.}{2018}]{Tonry2018}
{Tonry} J.~L.,  et~al., 2018, \mn@doi [\pasp] {10.1088/1538-3873/aabadf}, \href
  {https://ui.adsabs.harvard.edu/abs/2018PASP..130f4505T} {130, 064505}

\bibitem[\protect\citeauthoryear{{Xie} et~al.,}{{Xie} et~al.}{2025}]{Xie2025}
{Xie} C.,  et~al., 2025, \mn@doi [\nat] {10.1038/s41586-025-08920-4}, \href
  {https://ui.adsabs.harvard.edu/abs/2025Natur.641..608X} {641, 608}

\bibitem[\protect\citeauthoryear{{Zubko}, {Videen}, {Shkuratov}  \&
  {Hines}}{{Zubko} et~al.}{2017}]{Zubko2017}
{Zubko} E.,  {Videen} G.,  {Shkuratov} Y.,   {Hines} D.~C.,  2017, \mn@doi
  [\jqsrt] {10.1016/j.jqsrt.2017.07.026}, \href
  {https://ui.adsabs.harvard.edu/abs/2017JQSRT.202..104Z} {202, 104}

\makeatother
\end{thebibliography}




\renewcommand{\thefigure}{S\arabic{figure}}
\setcounter{figure}{0}
\renewcommand{\thetable}{S\arabic{table}}
\renewcommand{\theequation}{S\arabic{equation}}
\renewcommand{\thesection}{S\arabic{section}}
\setcounter{section}{0}
\cleardoublepage
\setcounter{page}{1}
\renewcommand\thepage{S\arabic{page}}
\section*{Supplemental Material}
\appendix
\renewcommand{\thefigure}{S\arabic{figure}}
\setcounter{figure}{0}
\renewcommand{\thetable}{S\arabic{table}}
\renewcommand{\theequation}{S\arabic{equation}}
\renewcommand{\thesection}{S}
\setcounter{section}{0}
\subsection{Observational details}
\label{obsdet}
\textit{Asteroid Terrestrial-impact Last Alert System Chile, Rio Hurtado:} The ATLAS telescope consists of a 0.5-m reflector telescope equipped with a charge-coupled device (CCD) camera, which has a pixel scale of 1.86 arcseconds per pixel. Survey observations use an ``orange'' o-band custom filter providing wavelength coverage between 560 nm and 820 nm, and an effective wavelength of 663 nm \citep[][]{Tonry2018}. W68 is located at El Sauce Observatory, Rio Hurtado.
\\
\textit{Kottamia Astronomical Observatory (KAO), Egypt, 1.88-m telescope/Kottamia Faint Imaging Spectro-Polarimeter (KFISP):} 
The KFISP is a combination direct imaging and spectrograph instrument mounted on the cassegrain focus of the KOA 1.88-m telescope \citep[][]{Osman2001,Azzam2022}. The KFISP in imaging mode consists of a 2048 pixel  x 2048 pixel array with a 8 arcmin x 8 arcmin field of view and a 0.243 arcsec/pixel pixel scale. The imager possesses Johnson-Cousins B,V, R, and I filters \citep[][]{Cousins1976}. In total, two 240 s B images, three 240 s V images, three 180 s R, and two 180 s I images were taken. The comet was tracked at a non-sidereal rate consistent with its skyplane motion over an airmass range of 1.50-1.56. The seeing measured in R-band images determined from full-width at half-maximum (FWHM) measurements of background stars of 1.33 arcsec during the 2025 July 2 observations. Kottamia Astronomical Observatory is located in the Cairo Governorate Desert, Egypt. 
\\
\noindent\textit{Palomar 200-inch telescope (P200)/Next Generation Palomar Spectrograph (NGPS)} The NGPS instrument is mounted at cassegrain focus on the P200 and possesses a guide camera with a 4 arcmin x 4 arcmin field of view, 0.26 arcsec/pixel spatial scale \citep[][]{Tang2023,Kasliwal2024}, and SDSS-equivalent g, r, and i filters \citep[$\mathrm{\lambda_{eff}}$ = 467.2 nm/614.1 nm/745.8 nm, FWHM = 126.3 nm/115.0 nm/68.3 nm;][]{Gunn1998}. The seeing in r-band was $\sim$1.6 arcsec measured by determining the full-width at half-maximum of background stars during the observations taken of \ti on 2025 July 3. Two g-band exposures and two i-band exposures were taken with a total cumulative exposure time of 150 s, and three r-band exposures were taken with a total cumulative exposure time of 270 s. The photometry from the g images was measured in individual images, and combined. Additionally, a sidereal-tracked image containing \ti was taken to provide star detections to compare with the surface brightness profile of \tins. The g, r, and i images were taken over an airmass range of 1.79 to 2.05 during the 2025 July 3 observations. The P200 is located on Palomar Mountain, California.
\\
\textit{Apache Point Observatory Astrophysical Research Consortium 3.5-m (ARC 3.5-m)/ Astrophysical Research Consortium Telescope Imaging Camera (ARCTIC):} ARCTIC has a 7.85 arcmin x 7.85 arcmin field of view and a 0.228 arcsec/pixel scale when binned in 2 x 2  mode \citep[][]{Huehnerhoff2016}. We conducted photometric time series observations of \ti using the ARCTIC imaging camera with the Sloan Digital Sky Survey (SDSS) g, r, i, and z filters with effective wavelengths 477 nm, 623 nm, 763 nm and 913 nm respectively \citep[][]{Fukugita1996}. In total, 5 x 120 s g band images,  6 x 120 s r band images, 5 x 120 s i band images, and 4 x 120 z band images were taken. The ARC 3.5-m was tracked at the rate of motion of the comet, the seeing was measured to be 1.47 arcsec, and the airmass range of the observations was 2.30 to 1.81 during the 2025 July 6 observations. The ARC 3.5-m is located at Apache Point, New Mexico.
\clearpage
\newpage
\begin{table}
\caption{\textbf{Observational details.}\label{t:obs}}
\begin{tabular}{llllllllll}
\hline
Facility & Instrument & UT Date and Time    & $\Delta^1$ & r$_H^2$ & $\alpha^3$     & Seeing$^4$                   & Airmass & Sky motion                 & m$^5$     \\
         &            &            & (au)     & (au)  & ($^{\circ}$) & (\arcsec) &         & (arcsec/h) & (mag) \\ \hline
W68      & ATLAS        & 2025-07-01 05:15 & 3.50    & 4.50 & 2.0          & 2.0                      & 1.05    & 73.3                      & 17.83  \\
KAO 1.88-m     & KFISP       & 2025-07-02 21:04 & 3.45    & 4.45 & 2.5         & 1.3                      & 1.53    & 74.1                     & 17.21   \\
P200     & NGPS       & 2025-07-03  07:27  & 3.43    & 4.43 & 2.6         & 1.6                      & 1.63    & 75.8                     & 17.86   \\
ARC 3.5-m     & ARCTIC       & 2025-07-06 03:23 & 3.36    & 4.34 & 3.6         & 1.5                      & 2.05    & 78.6                     & 17.77   \\
\end{tabular}
\begin{tablenotes}
\item \textbf{Notes.} (1) Geocentric distance, (2) heliocentric distance, (3) phase angle, (4) measured in science images, (5) o/R/r magnitude as measured in the images taken during the observations.
\end{tablenotes}
\end{table}

\begin{table}
\centering
\caption{\textbf{Orbital elements of \ti taken from JPL HORIZONS.} Solution access date on 2025 July 4 based on $\sim$319 observations submitted to the MPC between 2025 June 14 UTC and 2025 July 4. The osculating orbital elements are shown for the Julian date (JD) epoch 2,460,859.5. The 1~$\sigma$ uncertainties are given in parentheses.\label{tabel}}
\label{t:orbit}
\begin{tabular}{ll}
\hline
Heliocentric Elements&
\\ \hline
Epoch (JD) & 2,460,859.5\\
\hline
Time of perihelion, $T_p$ (JD) & 2,460,636.9175586$\pm$(3.76x10$^{-5}$)\\
Semi-major axis, $a$ (au) & --0.26549$\pm$(2.49x10$^{-3}$)\\
Eccentricity, $e$ & 6.0823$\pm$(9.76x10$^{-2}$)\\
Perihelion, $q$ (au) & 1.3493$\pm$(1.33x10$^{-2}$)\\
Inclination, $i$ ($^{\circ}$) & 175.10961$\pm$(4.87x10$^{-3}$)\\
Ascending node, $\Omega$ ($^{\circ}$) & 322.08$\pm$(0.11)\\
Argument of perihelion, $\omega$ ($^{\circ}$) & 128.07$\pm$(0.14)\\
Mean Anomaly, $M$ ($^{\circ}$) & --854.68$\pm$(10.65)\\
\hline
\end{tabular}
\end{table}

\begin{table}
\caption{\textbf{Properties of the known interstellar objects.}\label{t:objs}}
\begin{tabular}{lllllllllll}
\hline
name         & q    & e    & i      & g-r           & r-i           & r-z            & g-i           & Af$\rho$   & $\dot{M}$ & Refs. \\
             & (au) &      & (deg)  &               &               &                &               & (cm)       & (kg/s)    &       \\ \hline
1I/'Oumuamua & 0.26 & 1.20 & 122.74 & 0.84$\pm$0.05 & 0.36$\pm$0.1  & 0.36$\pm$0.10  & 1.20$\pm$0.11 & $\sim$0    & $\sim$0   & \citet[][]{Meech2017}       \\
2I/Borisov   & 2.01 & 3.36 & 44.05  & 0.63$\pm$0.05 & 0.20$\pm$0.02 & --0.03$\pm$0.11 & 0.83$\pm$0.05 & 143$\pm$10 & $\sim$1   & \citet[][]{Bolin20202I}       \\
3I/ATLAS     & 1.35 & 6.08 & 175.11 & 0.84$\pm$0.05 & 0.16$\pm$0.03 & 0.14$\pm$0.08  & 1.00$\pm$0.05 & 280.8$\pm$3.2  & $\sim$1   & This work.      
\end{tabular}
\begin{tablenotes}
\item \textbf{Notes.} (1) perihelion distance, (2) eccentricity, (3) inclination, (4) g--r colour, (5) r--i colour, (6) g--i colour, (7) Af$\rho$, (8) dust mass loss.
\end{tablenotes}
\end{table}

\bsp	
\label{lastpage}
\end{document}